\documentclass[pre,amsmath,amssymb,showpacs,floats]{revtex4}

\usepackage{graphicx}

\def \beq{\begin{equation}}
\def \eeq{\end{equation}}
\def \bea{\begin{eqnarray}}
\def \eea{\end{eqnarray}}
\def \karman{$\mathrm{K\acute{a}rm\acute{a}n}$}
\def\integral{\int}
\def\proportionalto{\propto} 
\def \half {\frac 12} 
\def \muchlessthan {\ll}
\def\aboutequal{\simeq} 
\def\definedas{\equiv}
\def\lessthanorabout {\mathrel{\raise.3ex\hbox{$<$\kern-.75em\lower1ex\hbox{$\sim$}}}}
\def\greaterthanorabout {\mathrel{\raise.3ex\hbox{$>$\kern-.75em\lower1ex\hbox{$\sim$}}}} 
\def\goesto{\to}
\def\goesas{\sim} \def\infinity{\infty}

\begin{document}

\title{Crescent Singularities in Crumpled Sheets}
\author{Tao Liang and Thomas A. Witten}
\affiliation{The James Franck Institute and the Department of Physics, The University of Chicago, 5640 S. Ellis Avenue, Chicago, IL 60637}

\date{\today}


\begin{abstract}
We examine the crescent singularity of a developable cone in a setting similar to that studied by Cerda \textit{et al} [Nature \textbf{401}, 46 (1999)]. Stretching is localized in a core region near the pushing tip and bending dominates the outer region. Two types of stresses in the outer region are identified and shown to scale differently with the distance to the tip. Energies of the $d$-cone are estimated and the conditions for the scaling of core region size $R_c$ are discussed. Tests of the pushing force equation and direct geometrical measurements provide numerical evidence that core size scales as $R_c \sim h^{1/3} R^{2/3}$, where $h$ is the thickness of sheet and $R$ is the supporting container radius, in agreement with the proposition of Cerda \textit{et al}. We give arguments that this observed scaling law should not represent the asymptotic behavior. Other properties are also studied and tested numerically, consistent with our analysis.   

\end{abstract}

\maketitle

\section{INTRODUCTION}

As we crumple a piece of paper in our hands, two types of singular structures appear in the crumpled paper: folding ridges and point-like vertices. Energies are condensed into a network of such singularities. The properties of ridges have been studied thoroughly. Scaling laws governing the energy and size of the ridge have been obtained analytically and tested numerically [6-9]. Point-like singularities are also studied extensively [1-4, 11, 12, 16, 17], however, current understanding of their properties is not as complete as that of ridges.  

\begin{figure}[thb]
\begin{center}
\includegraphics[width=0.45\textwidth]{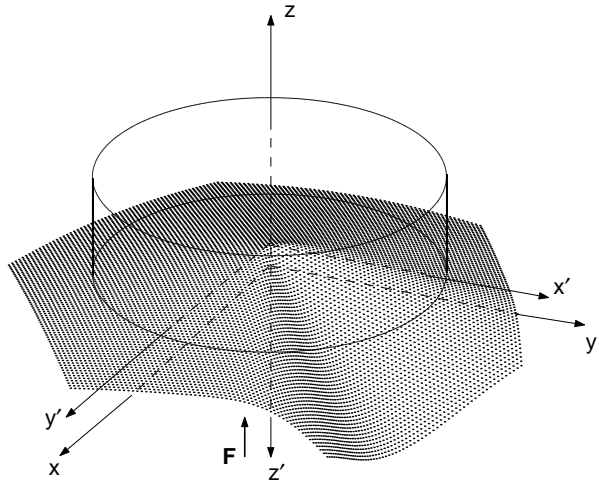}
\caption{A $d$-cone appears when we push the center of an originally flat sheet against the edge of a cylindrical container from below with force $F$. This is a typical simulated $d$-cone shape, with side length $l=60a$, container radius $R=38a$, displacement $d=0.15R$ and thickness $h=0.102a$, where $a$ is the lattice spacing. Two sets of coordinate systems are shown. }
\label{shape}
\end{center}
\end{figure}

\begin{figure}[!hbt]
\begin{center}
\includegraphics[width=0.45\textwidth]{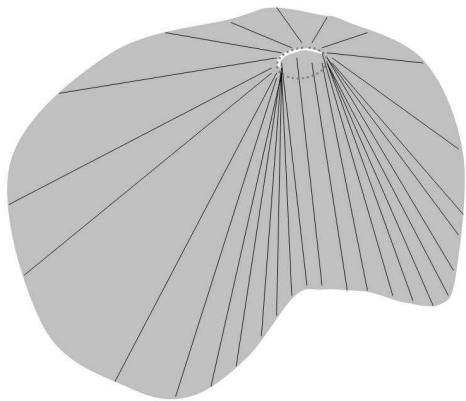}
\caption{Sketch of the core region of a $d$-cone. Crescent area is shown in white. Directors are shown as black lines. Many directors converge toward the two tips of the crescent. Dashed circle tangent to the crescent at its center defines the radius $R_c$.}
\label{crescent}
\end{center}
\end{figure}

In this paper, we consider a single conical vertex studied by Cerda \textit{et al} [1, 2]. One experimental realization is to push the center of a circular elastic sheet of radius $R_p$ axially into a cylindrical container of radius $R$, as illustrated in FIG.~\ref{shape}. This is the simplest volume-restricting deformation of the sheet and causes the center of the sheet to move into the container by a distance $d$. It is useful to express the deflection of the sheet by $\epsilon \equiv d/R$. Due to the constraint of unstretchability, the sheet deforms into a non-axisymmetric conical surface which is only in partial contact with the edge of the container. In the limit that thickness $h$ of the sheet goes to zero, since bending modulus ($\sim h^3$) vanishes faster than the stretching modulus ($\sim h$), there would be pure bending over the sheet and Gaussian curvature would be zero everywhere. Mathematically, such a conical surface is called perfectly developable cone \cite{love} ($d$-cone). In this limit, some models about the shape of the $d$-cone have been proposed [2-4, 17]. These models only give outer region solutions of $d$-cone shape, in the sense that they do not consider the stretching energy that is inevitable on a real sheet with finite thickness. For a real sheet, it must stretch near the tip, because otherwise, the curvature at the tip would be divergent, since curvature goes as $1/r$, where $r$ is the distance to the tip, thus causing divergent energy. Therefore, it is the finite thickness that causes the sheet to stretch greatly in a small region near the tip. This small region is called the core region. It is where energetically expensive stretching is localized and its size is governed by the competition of the bending and stretching energies of the whole cone. As a result of the stress focusing in the core region, crescent-like shapes come out where bending stresses are big, as shown in FIG.~\ref{crescent}. In addition, as we will see later, finite thickness also causes a small amount of strains in the outer region.   

One problem of great interest is concerned with the size of the core region, characterized by the radius of curvature of the crescents. We want to know whether there is scaling behavior of core size and to determine any scaling exponents. Intuitively, we expect that this size $R_c$ should have a dependence on thickness $h$, since $R_c$ goes to zero as $h$ goes to zero. Cerda \textit{et al} \cite{cerda-nature} propose that $R_c$ scales as $R_c \sim h^{1/3} R^{2/3}$. This says that besides $h$, the supporting container radius $R$ also determines $R_c$. This is surprising because stretching energies are supposed to be localized in the core region, so the core should not be able to know about the length of outside container radius. In this work we explore this problem as well as other properties of a $d$-cone. In Sec.~II, we describe the energies and forces that give rise to the crescent singularity. We first study the elastic properties of a truncated $d$-cone formed by cutting a cap region at the center of a regular $d$-cone, and then investigate the energetic variations as we join the cap region into the truncated $d$-cone. From them, we discuss the conditions for the existence of scaling behavior of $R_c$. Details of numerical models of simulating an elastic sheet and producing desired shapes are presented in Sec.~III. After numerical results that agree with the predicted $d$-cone properties are shown, we use two different methods to look for scaling exponents of $R_c$. Finally, the limitations of and implications from our findings and future work are discussed in Sec.~IV. 

\section{THEORIES}

We begin by stating the connection between the deformation of the sheet and its elastic energy. In a thin $d$-cone the dominant energy is due to the bending distortion in outer region far from the crescent. We discuss the form of this energy and the stresses associated with it. Next we outline the energetic effects of the crescent region. Finally, we focus on how each of these energies is influenced by a change in the crescent radius $R_c$. We sketch how scaling properties of the energy produce scaling behavior in the sheet.

We first specify two coordinate systems. We define $x-y$ as the horizontal plane of the supporting cylindrical container's edge, with origin on the axis of cylinder and $z$ axis pointing upward (see FIG.~\ref{shape}). We then define an $x'-y'$ plane as parallel to $x-y$ plane but with origin at the tip of the sheet and $z'$ axis pointing opposite to $z$ axis. $x'-y'$ plane is displaced from the $x-y$ plane by a distance $d$. Next we define $(\rho, \theta)$ as the polar coordinates in the $x'-y'$ plane. In the cylindrical coordinates $\{\rho \theta z'\}$, $d$-cone is described by $z'=\rho \psi(\theta)$, in the limit that thickness $h \rightarrow 0$. We shall denote this limiting surface as the \textit{asymptotic $d$-cone}. 

In equilibrium, the sheet assumes a conformation that minimizes the elastic energy. Thus, the actual core radius is that which minimizes this energy. Two forms of energy must be considered, bending energy $B$ and stretching energy $S$. The bending energy density is proportional to the square of the total curvature $C(r)$ \cite{mansfield}. This $C(r)$ defined as the trace of the curvature tensor, is sometimes called the mean curvature \cite{nelson}. The constant of proportionality is called the bending modulus $\kappa$. Thus
\beq
B = \half \kappa \integral d A ~C(r)^{2} ~~, 
\eeq
where $\integral d A$ denotes the integral over the surface. In general there is a second form of bending energy, proportional to the average Gaussian curvature. We show later in this section that this Gaussian curvature energy is unimportant for the present system.  

The stretching energy density is proportional to the square of the strain tensor $\gamma$. For an isotropic material, there are two forms of stretching energy with constants of proportionality $Y h$ and $Y_1 h$ \cite{landau}
\beq 
S = \half \integral d A \left [Y h~(Tr ~\gamma)^2 + Y_1 h~ (Det ~\gamma) \right]~~,
\eeq
where $Y$ is Young's modulus, and $Y_1$ is a convenient combination of $Y$ and dimensionless Poisson's ratio $\nu$. They are related to bending modulus through $\kappa = Y h^3/12(1-\nu^2)$.

The actual strain and curvature fields are those which minimize the $B + S$. The variational minimization amounts to a statement that the normal forces on each element must balance. This statement is known as the ``force von \karman~equation" 
\beq
\partial_{\alpha} \partial_{\beta} M_{\alpha \beta} = \sigma_{\alpha \beta} C_{\alpha \beta}+ P~~,
\label{forcevon}
\eeq
where $M_{\alpha \beta}$ are torques per unit length, $\sigma_{\alpha \beta}$ are in-plane stresses, and $P$ is external force per unit area of the sheet.  

When we push a circular sheet of radius $R_p$ into a container to form a $d$-cone, strain and curvature distortions are created and elastic energy is stored. This energy arises from several effects. It is convenient to discuss it by creating the final surface in two stages. In the first stage, we assume a value for $R_c$ and cut a circular hole of that radius in the center of the sheet. We then force this perforated sheet into the container by exerting tension on the inner edge of the hole. In the unstretchable limit, $h \muchlessthan R$, this shape becomes a truncated $d$-cone, which serves as the outer region of a regular $d$-cone. In this region, the curvature at every point vanishes in the radial direction. The total curvature $C$ has the form $C = \phi(\theta)/r$. Its bending energy $B_0$ is thus given by 
\beq
B_0 = \half \kappa \integral_{R_c}^{R_p} r dr/r^2 \integral d\theta \phi^2(\theta) \proportionalto \kappa \log(R/R_c)~~.
\eeq
In the limit $h \muchlessthan R_c \muchlessthan R_p$, this $B_0$ dominates the energy, as we justify below. Then the function $\phi$ takes the form that minimizes $B_0$ subject to the constraints on the $d$-cone. In this geometry, the force von \karman~equation simplifies greatly, and the form of $\phi(\theta)$ as well as the associated stresses can be found explicitly \cite{cerda-prl}\cite{maha-new}.   

We first determine the transverse stress $\sigma_{\theta \theta}$ using Eq. (\ref{forcevon}). The nonzero components of torque tensor $M$ follow from the constitutive law implicit in the energy equations \cite{mansfield}: $M_{rr}=\kappa \nu C_{\theta \theta} = \kappa \nu \phi(\theta)/r$ and $M_{\theta \theta}= \kappa C_{\theta \theta} = \kappa \phi(\theta)/r$. The force von \karman~equation then reduces to $\partial^2_r M_{rr} + \partial^2_{\theta} M_{\theta \theta}/r^2 = \sigma_{\theta \theta} C_{\theta \theta} + P$, from which we obtain 
\beq
\sigma_{\theta \theta} = \frac{\kappa}{r^2}\left(2\nu + \frac{\ddot{\phi}(\theta)}{\phi(\theta)} \right) - \frac{Pr}{\phi(\theta)} ~~,
\label{strain}
\eeq
where the dots above functions denote $\theta$ derivatives. In the same region where only $C_{\theta\theta}$ is nonzero, there is no pressure from external force acting on the sheet, i.e. $P=0$. Therefore, according to Eq. (\ref{strain}), stress $\sigma_{\theta \theta}$ goes as $\kappa/r^2$. These stresses arise from the requirement that the normal force due to changing torques is balanced by the normal force due to in-plane tension of the sheet. We call them type I stresses and denote them by symbol $\sigma^{(1)}$.

On the other hand, we consider the force balance of a region that encloses area between inner radius $R_c$ and outer radius $r$, where $r$ can take values between $R_c$ and $R$. The tension exerted on the inner edge of this region is equivalent to the central pushing force of a regular $d$-cone. This force must be balanced by the force due to radial in-plane stress $\sigma_{rr}$ on the outer perimeter of the region. Let $\beta$ be the angle between a radial line or generator of the $d$-cone and the horizontal. Since $\tan{\beta}=\psi(\theta)$, we have $\sin{\beta} = \psi(\theta)/\sqrt{1+\psi^2 (\theta)}$. The balance of vertical forces yields
\beq
\int \sigma_{rr}  r \sin{\beta} d \theta = F ~~,
\label{above}
\eeq
which holds for every value of $r$ from $R_c$ to $R$. It is easy to see that type I stresses alone can't satisfy this equation, since type I stresses go as $1/r^2$, they would give a $1/r$ prefactor on the left side of equation, while the right-side of equation is independent of $r$. Therefore, the integral from type I stresses must vanish and there must exist some additional stresses in the outer region that scale as $1/r$ to satisfy Eq. (\ref{above}). We call these stresses as type II stresses and denote them by $\sigma^{(2)}$. They persist up to the supporting container edge, where normal force from the container counteracts the external pushing force. We can write $\sigma^{(2)} r = F e(\theta)$, where $e(\theta)$ is a function only of $\theta$ and satisfies $\int e(\theta) \sin{\beta} d \theta = 1$. It is obvious that $e(\theta)$ is of order unity. 

To estimate the magnitude of type II stresses, we notice that $F=\partial E/\partial d=(\partial E/ \partial \epsilon)/R \approx \kappa/R$. Thus $\sigma^{(2)} \approx F/r \approx \kappa/(rR)$. The type II stresses are comparable with type I stresses only near the container edge since the ratio $\sigma^{(1)} / \sigma^{(2)} \approx r/R$ for $R_c<r<R$. It is noted that they are both due to nonzero thicknesses. Since in-plane stress tensors are related to strain tensors through \cite{landau} $\sigma_{\alpha \beta}=[Yh/(1-\nu^2)](\gamma_{\alpha \beta} + \nu \epsilon_{\alpha \rho} \epsilon_{\beta \tau} \gamma_{\rho \tau})$, we have strains $\gamma \approx \sigma /(Yh) $. The stretching energy of truncated $d$-cone is then 
\beq
S_0 \approx \kappa h^{-2} \int \left((\sigma^{(1)} + \sigma^{(2)})/(Yh)\right)^2 r dr d \theta \approx \kappa h^2/R_c^2~~.
\label{outs}
\eeq

We now examine the profile of external force along the container edge. Here the external normal force pressure $P$ of Eq. (\ref{forcevon}) is nonvanishing. This  normal force causes a small transverse deflection of the sheet and hence induces both curvature and strain near the edge. Noticing that the curvature induced is in radial direction and has opposite sign with $C_{\theta\theta}$, we expect mean curvature to be reduced near the edge. Due to the translational symmetry along the region of the surface in contact with the cylindrical container, the normal force pressure is found to be independent of azimuthal angle $\theta$. However, besides this $\theta$-independent term, Cerda and Mahadevan \cite{maha-new} show that a $\delta$-function term of $\theta$ emerges in the normal force pressure expression in order to make the torque balance, and this singularity happens at the take-off angular positions, where the sheet begins to bend away from the container. Since stresses and curvatures should avoid singular behavior for the sake of energy, according to Eq. (\ref{strain}), there must exist a $\delta$-function term in $\ddot{\phi}(\theta)$ to cancel that from normal force pressure $P$. Integrating over $\theta$, we conclude that $\dot{\phi}(\theta)$ should have a jump at the take-off positions. This result is consistent with the geometrical requirement. More quantitatively, following Ref. \cite{maha-new}, we find that the ratio of the normal force contributed by the $\delta$-function term to that contributed by the $\theta$-independent term is $\tan{\theta_c}/[2(\pi -  \theta_c)]$, where $ \theta_c$ is the half aperture angle of non-contact region. For small deflections, taking $\theta_c \approx 1.21$ rad \cite{cerda-prl}, we obtain the value of this ratio 0.69. We will numerically verify this ratio later.

Having discussed stresses and forces, we now consider in detail the bending energy $B_0$. Bending energy comes from both total curvature and Gaussian curvature. Since the truncated $d$-cone has no Gaussian curvature, we only need to take account of the total curvature contribution. The reduced curvature $\phi(\theta)$ is related to the reduced height $\psi(\theta)\definedas z'/\rho$ defined in the beginning of this section: $\phi(\theta) = \psi(\theta)+\ddot{\psi}(\theta) - \psi(\theta) \dot{\psi}^2(\theta)/(1+\psi^2(\theta))$. For small to moderate deformations of the sheet ($\epsilon \lessthanorabout 0.4$), $C \approx (\psi+\ddot{\psi})/r \aboutequal \epsilon/r$. Throughout the paper we shall focus on this moderate range of $\epsilon$. The bending energy $B_0$ has the form
\beq
B_0 = \frac{\kappa}{2} \int C^2 dA \approx G_1 \kappa \epsilon^2 \ln{(R_p/R_c)}~~,
\label{outerbending}
\eeq
where $G_1$ is a geometrical factor. Comparing this with $S_0$ from Eq. (\ref{outs}), we see that bending dominates the outer region.

Although Gaussian curvature is zero everywhere in a truncated $d$-cone, it is certainly not the case for a regular $d$-cone. The Gaussian curvature contribution to the bending energy is $B_{G} = (\kappa_G/2) \int K d A$, where $\kappa_G$ is Gaussian curvature coefficient, $K$ is Gaussian curvature and $A$ is the area. According to the Gauss-Bonnet theorem, the integral of Gaussian curvature over a region $M$ of a surface is related \cite{love} to the integral of the geodesic curvature $\kappa_g$ over the boundary of that region through $\int_M K dA = 2\pi - \int_{\partial M} \kappa_{g} d s$. Choosing $M$ to be the whole sheet without perforation, we get $B_{G} = (\kappa_G/2) (2\pi - \int \kappa_{g} ds)$. If the sheet were a perfectly developable conical surface, the geodesic curvature at the perimeter would be the same as that of a regular cone of same size: $\kappa_g = 1/R_p$, where $R_p$ is the perimeter-to-tip distance. However, due to stretching of the real sheet and the pushing of container edge, the shape is not perfectly developable so there are variations in the perimeter-to-tip distances along the boundary, which make $\kappa_g$ deviate from $1/R_p$. Since stretching is small compared with bending and the sheet is nearly developable over most of the surface, this deviation is small compared with $1/R_p$, and it should depend on thickness $h$. We suppose that $\kappa_g$ is a regular function of $h$, and may be expressed $\kappa_g \approx (1/R_p) \left( 1 + (h/R_p) \Theta (\theta,\epsilon) \right)$, where $\Theta$ is a dimensionless function of $\theta$ and $\epsilon$. Substituting this into the equation for bending energy, we obtain $B_G = - (\kappa_G/2) (h/R_p) \int \Theta(\theta, \epsilon) d \theta$. Compared with Eq. (\ref{outerbending}), this energy is only a small fraction $(h/R_p)/\ln(R_p/R_c)$ of the bending energy contributed by total curvature. 

\begin{figure}[!hbt]
\begin{center}
\includegraphics[width=0.45\textwidth]{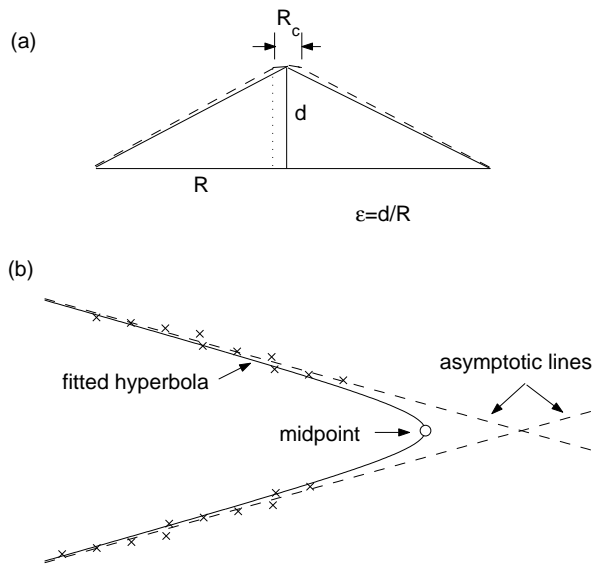}
\caption{(a) Sketch of a cross-section of the $d$-cone. Solid hypotenuse is for the $d$-cone when $h \rightarrow 0$ and has slope $\epsilon$; dashed hypotenuse is for a real sheet ($h \ne 0$) and has slope $\epsilon'$. (b) Sketch of fitting a hyperbola to find $R_c$ from geometry.}
\label{two_sketch}
\end{center}
\end{figure}

The truncated $d$-cone discussed above has no crescent singularity. This singularity arises from the further constraints of filling in the hole in the truncated cone. The flat disk of radius $R_c$ that was removed must be distorted in order to join onto the truncated cone. This distorted cap exerts forces on the truncated $d$-cone and distorts it in turn. All of these distortions add elastic energy to the energy $B_0 + S_0$ of the initial truncated $d$-cone. One obvious addition is the bending energy $B_G$ due to Gaussian curvature discussed above. Another simple consequence of adding the core region is to alter the slope of the outer region generator from $\epsilon$ to a higher value $\epsilon'$ as shown in FIG.~\ref{two_sketch}(a). To relate $\epsilon'$ to $\epsilon$, we imagine a cross-section of the $d$-cone extending from the container edge to the tip and back down to the edge (see FIG.~\ref{two_sketch}(a)). In the truncated $d$-cone, the generators have the same slope $\epsilon$ as the straight line from the container edge to the vertex, as shown by the solid lines in FIG.~\ref{two_sketch}(a). Now for the real sheet, the cap is curved, with a curvature radius of order $R_c$, but still, its uppermost point remains at height $d$. This rounding at the top necessarily increases the slope of the outer cone generator, as illustrated by the dashed lines in the same figure. When deformation is small, $\epsilon'\approx d/(R-R_c/2) \approx \epsilon (1+R_c/(2R))$. The altered bending energy is given by Eq. (\ref{outerbending}) with $\epsilon$ replaced by $\epsilon'$: $B'_0 \approx G_1 \kappa \epsilon^2 (1+R_c/R) \ln (R_p/R_c)$. This increases the $B_0$ energy by a factor $(1 + R_c/R)$. This addition to the energy favors small $R_c$.

Besides these variations, there are other forms of added energy. We denote them as $\Delta E$. It consists of stretching energy $\Delta E_0$ added to the original region plus the energy $E_c$ of the cap or core region. The added energy $\Delta E$ is expected to be much smaller than the dominant energy $B_0$ discussed above. Nevertheless, $\Delta E$ is all-important in determining the core-radius $R_c$.

Further distortions increase the energy of the outer region. The crescent singularity increases the curvature near the tips of the crescent. As shown in FIG.~\ref{crescent}, the generators, initially distributed smoothly around the inner hole, are now concentrated at the tips of the crescent, increasing the curvature there. The generators splay outward from these tips. Observation suggests that a finite fraction of the generators are pushed to the tips. These generators appear to splay outward less rapidly when the size $R$ is increased. These observations suggest that a) the crescent increases the exterior curvature energy, and b) this energy decreases as the size $R \goesto \infinity$ for a given $R_c$. This energy penalty favors small $R_c/R$. We do not have a more specific estimate for this energy at present.

We now consider the energy $E_c$ of the cap or core region itself. In order to bridge the truncated cone, the average curvature $C$ must be of order $\epsilon /R_c$. If the curvatures are uniform, the associated bending energy is of order $\kappa \integral dA C^2 \aboutequal \kappa \epsilon^2 $. However, such uniform curvatures are not optimal. The smoothly-curved cap region has typical Gaussian curvatures of order $\epsilon^2/R_c^2$, consisting of a negative part in the buckled region and a comparable positive part elsewhere. The presence of Gaussian curvature induces strain of order $\epsilon^2$, and a stretching energy of order $\kappa h^{-2} \epsilon^4 R_c^2$.  The system may alleviate this large stress energy by concentrating the curvature, as in the simple stretching ridge \cite{alex3}. The observed crescent singularity is presumably the result of this concentrated curvature. The length of this crescent is of order $R_c$; its width $w$ is evidently that which minimizes $E_c$.  If $w$ is a fixed fraction of $R_c$, the above estimates show that $E_c$ grows as $R_c^2$. To improve this energy, $w/R_c$ must go to zero as $R_c$ grows. The total curvature is then of order $1/w$, and the bending energy of the core goes as $\kappa w R_c/w^2 \goesas R_c / w$. This energy must grow indefinitely as $R_c \goesto \infinity$. Evidently, the core energy favors small $R_c$.

With these energies in mind, we now investigate how $R_c$ is determined. It is clear that $R_c$ is controlled by the competition of energies. The total energy is $E = S_0 + B'_0 + B_G + \Delta E$. Our purpose is to find an optimal $R_c$ that minimizes the total energy. First, taking derivative of $B'_0$ with respect to $R_c$, we have $\partial B'_0 / \partial R_c = G_1 \kappa \epsilon^2 [-1/R_c + (\ln(R_p/R_c)-1)/R ]$. In the regime of interest, $R$ is comparable with $R_p$, and $R_c \ll R$, so the first term dominates the second term: $\partial B'_0 / \partial R_c \approx - G_1 \kappa \epsilon^2 /R_c$. This implies that $B_0$ favors large $R_c$. Next, we notice that since $\partial B_G / \partial R_c$ and $\partial S_0 / \partial R_c$ are much smaller than $\partial B'_0 / \partial R_c$ in this regime, the competition happens mainly between outer bending energy $B'_0$ and the added energy $\Delta E$. Though we know little about the form of the added energy $\Delta E$, the above observations suggest that a) it favors small $R_c$, b) it decreases as $R$ increases, and c) it involves stretching, and thus increases with decreasing thickness $h$. To illustrate the possible effect of this energy, we posit that $\Delta E$ is homogeneous in $R_c$, $R$, and $h$
\beq
\Delta E =  \kappa g(\epsilon) R_c^s ~R^{-t} ~ h^{-s+t} ~~,
\label{assumption}
\eeq
where $g(\epsilon)$ is an unknown function of $\epsilon$ and $s$ and $t$ are unknown positive exponents. The $h$ power law is determined from $s$ and $t$ by dimensional consistency. The energy minimization with respect to the variational parameter $R_c$ then implies 
\beq
0 = R_c \frac{\partial E}{\partial R_c} = -G_1 \kappa \epsilon^2 + s \Delta E ~~.
\label{B0-delta-E-balance}
\eeq
Using the assumed power-law form of $\Delta E$, we infer
\beq
R_c \goesas R^{t/s} h^{1-t/s}~~.
\eeq
Thus the assumption of Eq. (\ref{assumption}) implies that $R_c$ should increase as a power of $R$. Since $R_c$ cannot exceed $R$, the power must be smaller than unity, so that $t$ must be smaller than $s$.

Though this assumed form for $\Delta E$ gives a pleasing result, we warn that it cannot be correct in the limit of large $R/h$. If it were correct, it would imply that $R_c/h$ goes to infinity with $R/h$. This implies that the core energy $E_c$ must go to infinity relative to $\kappa \epsilon^2$, as shown above. The increase of $E_c$ occurs irrespective of $R$. Thus $E_c$ must grow to dominate the total energy. Such a large $E_c$ is not compatible with a minimum of total energy. It must be balanced by some other energy favoring large $R_c$. No such energy is apparent. Thus $R_c$ cannot grow indefinitely. Conversely, any observed power-law growth of $R_c$ must cease for sufficiently large $R/h$. In the numerical section, we will give an empirical discussion of the range of $R/h$ in which the scaling of $R_c$ holds.

These conclusions contrast with the scaling argument proposed in the initial studies of the d-cone \cite{cerda-nature}\cite{maha-new}. Thus it is important to revisit this scaling argument and show how the contradiction with our result arises. The authors attribute the scaling to an energy balance between bending and stretching in the core region. To estimate the stretching energy, they consider a director traversing the sheet from one edge, through the core and on to the opposite edge, such as the dashed line in FIG. \ref{two_sketch}(a). They note that increased $R_c$ increases the length of this chord (relative to the solid line) by a fraction of order $(R_c/R)^2$. If this strain is presumed uniform, the associated stretching energy in the core is of order $\kappa h^{-2} R_c^2 (R_c/R)^4$. They balance this energy against the bending energy in the core, presumed to be of order $\kappa \epsilon^2$. This balance yields $R_c \goesas R^{2/3} h^{1/3}$.

This argument seems questionable, since it ignores energies much larger than the ones it includes. Its assumption of uniform strain leads to an stretching energy of uniform density outside as well as within the core. The argument includes that part of this energy lying within the core while ignoring the much larger part outside the core. If one tries to repair the estimate by supposing the excess length resides only in the core, then the strain becomes independent of $R$ and the $R_c$ must be independent of $R$. In addition the argument ignores the large stretching energy in the core arising from the necessary Gaussian curvature within the core itself, as discussed above. Likewise, it takes no account of the strong difference between the two principal curvatures in the crescent.

Our analysis of the energy given above allows us to infer the scaling of $R_c$ based on measurements of the central pushing force, $F$. The inference is valid so long as the dominant energy is the $B_0$ energy above. The energy balance equation (\ref{B0-delta-E-balance}) allows us to write $\Delta E = G_2 \kappa \epsilon^2$, where $G_2$ is a numerical constant. Thus
\beq
E \approx G_1 \kappa \epsilon^2 ~\ln (R_p/R_c)  +  G_2 \kappa \epsilon^2~~.
\label{energy}
\eeq
The pushing force follows from $F=\partial E/\partial d=(\partial E/ \partial \epsilon)/R$. To explore the scaling of $R_c$, we write $R_c$ as $R_c = h^p R^q R_p^{\lambda} A(\epsilon)$, where $p$, $q$ and $\lambda$ are scaling exponents to be determined, and $A(\epsilon)$ is a function only of $\epsilon$. Dimensional consistency requires that $p+q+\lambda=1$. Previous work \cite{cerda-nature}\cite{cerda-prl} did not consider the possibility that $R_c$ has a dependence on $R_p$. Here we address it by introducing $\lambda$. From Eq. (\ref{energy}) and the scaling expression of $R_c$, the pushing force is calculated to be 
\beq 
\label{forceeq}
F \approx \frac{2G_1\kappa\epsilon}{R}\left(-p\ln{h}-q\ln{R}+(1-\lambda)\ln{R_p}+f(\epsilon)\right)~~,
\eeq
where $f(\epsilon) = -\ln{A(\epsilon)}+G_2/G_1-\epsilon \dot{A}(\epsilon)/2A(\epsilon)$ is a function only of $\epsilon$. We will use this equation to test the scaling relations in the numerical section, which follows below.

\section{NUMERICS AND FINDINGS}

\subsection{Numerical Model}

We model an elastic sheet by a triangular lattice of springs of un-stretched length $a$ and spring constant $k$ after Seung and Nelson \cite{nelson}. Bending rigidity is introduced by assigning an energy of $J(1-\hat{n_1}\cdot \hat{n_2})$ to every pair of adjacent triangles with normals $\hat{n_1}$ and $\hat{n_2}$. When strains are small compared to unity and radii of curvature are large compared to the lattice spacing $a$, this model is equivalent to an elastic sheet of thickness $h=a\sqrt{8J/k}$ made of an isotropic, homogeneous material with bending modulus $\kappa=J\sqrt{3}/2$, Young's modulus $Y=2ka/h\sqrt{3}$ and Poisson's ratio $\nu=1/3$. Lattice spacing $a$ is set to be 1. The shape of the sheet in our simulation is a regular hexagon of side length $R_p$. The typical value of $R_p$ is $60a$. 

To obtain a single $d$-cone shape, we need to simulate the constraining container edge and pushing force. As shown in FIG.~\ref{shape}, the edge lies in the $x-y$ plane and is described by equation $x^2+y^2=R^2$. Pushing in the center of the sheet is accomplished by introducing a repulsive potential of the form $U_{\mathrm{force}}(z_1)=-Fz_1$, where $z_1$ is the $z$ coordinate of the lattice point in the center and $F$ is the magnitude of the pushing force. This force is applied in the positive $z$ direction. The constraining edge is implemented by a potential of the form $U_{\mathrm{edge}}=\sum C_p H(z_i)/(((\sqrt{x_i^2+y_i^2}-R)^2+z_i^2)^4+\xi^8)$, where $\xi$, $C_p$ are constants and the summation is over all lattice pionts with coordinates $(x_i, y_i, z_i)$. $H(z)$ is the unit step function, which makes certain that this potential only acts on the lattice points that have already moved into the container (those with $z_i > 0$). The force associated with it decays rapidly once the lattice points go away from the edge. The conjugate gradient algorithm \cite{recipe} is used to minimize the total elastic and potential energy of the system as a function of the coordinates of all lattice points. FIG.~\ref{shape} shows one such minimized configuration of the lattice grid. The simulated sheet is indeed only in partial contact with the container edge. In the rest of this section, we shall first compare our results to the $d$-cone predictions of previous work. The good agreement indicates that our numerical realization is reliable. Next we shall investigate the scaling behaviour of $R_c$.  

\subsection{Curvature Profile}

\begin{figure}[!hbt]
\begin{center}
\includegraphics[width=0.5\textwidth]{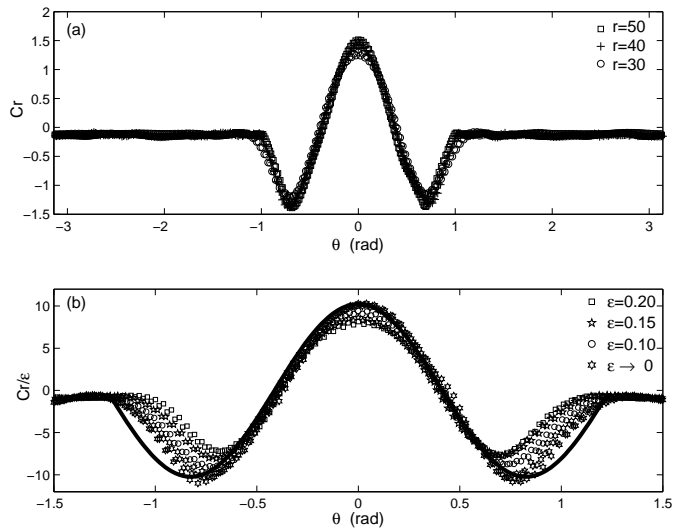}
\caption{(a) Azimuthal profile of normalized total curvature on a $d$-cone at three fixed distances from the tip. Curvature is positive for the concave region; negative for the convex region. Curvature is normalized by $1/r$, where $r$ is the fixed distance from the tip. This plot is for the $d$-cone shape shown in FIG.~\ref{shape} at distances $30a$ (circle), $40a$ (plus) and $50a$ (square) from the tip. The container touches the $d$-cone surface at $r=R\sqrt{1+\epsilon^2} \approx 38.4a$. (b) Normalized curvature profiles for four different values of $\epsilon$ at fixed thickness $h=0.102a$. Curvature is normalized by $\epsilon/r$. Profile at $\epsilon \rightarrow 0$ is obtained by extrapolating from profiles at $\epsilon=0.10$, 0.15 and 0.20. Solid line is the exact solution as both $h \rightarrow 0$ and $\epsilon \rightarrow 0$ \cite{cerda-prl} \cite{maha-new}.}
\label{curvature}
\end{center}
\end{figure}

We first test our simulation by measuring the curvature profile of the sheet, and comparing it to the prediction under certain limiting situations. We determine the curvatures approximately from each triangle in the sheet. For this measurement, we take the curvature tensor to be constant across each triangle. We calculate it using the relative heights of the six vertices of the three triangles that share sides with the given triangle \cite{brian}. The six relative heights $w_i$ normal to the triangle surface are fit to a function of the form
\beq
w_i=b_1+b_2 u_i+b_3 v_i +b_4u_i^2 +b_5 u_i v_i+b_6 v_i^2, ~~i=1, \ldots, 6
\eeq
where \{$u_i,v_i,w_i$\} are coordinates of the vertices in a local coordinate system that has $w$ axis perpendicular to the surface of the given triangle. This choice of local coordinate system ensures that $b_2$ and $b_3$ are negligible so that curvature tensors can be determined only from the coefficients of quadratic terms. In practice, our numerical findings do show that the values of $b_2$ and $b_3$ are on the order of $10^{-2}$ or lower. Therefore, curvature tensors follow immediately from the identification $C_{uu}=2 \times b_4, C_{vv}=2 \times b_6, C_{uv}=b_5$. Total curvature $C$ is defined as the trace of curvature tensor: $C=C_{uu}+C_{vv}$. FIG.~\ref{curvature}(a) gives a typical plot of total curvature versus azimuthal angle on a $d$-cone surface at three different distances from the tip. Curvature is measured in units of $1/r$ and taken to be negative in the convex region, where the sheet touches the supporting container. The three curves mostly collapse into one single curve. This roughly verifies that curvature goes as $1/r$. However, near $\theta=0$, the normalized curvature becomes systematically smaller for smaller $r$. We attribute this departure to the influence of the stretched crescent region.

The shapes of the curves are in qualitative agreement with our analysis below. The surface is in contact with the container for angles $\theta$ larger than some $\theta_c$ in magnitude. For $|\theta| > \theta_c$, the normalized curvature at all $r$ shown is constant and its value is that of a simple cone of the same $\epsilon$. As $|\theta|$ becomes less than the ``take-off'' angle $\theta_c$, the surface curves inwardly away from the container and the total curvature begins to increase in negative direction. The crescent singularity is most likely to live near the region where total curvature reaches its negative maximum. As $|\theta|$ decreases further, the curvature goes to zero, changes signs, and reaches a central positive maximum at $\theta=0$, which corresponds to where the sheet is maximally deflected.

As thickness $h$ and deformation $\epsilon$ go to zero, Cerda and Mahadevan obtained the exact solution of curvature profile \cite{cerda-prl}\cite{maha-new}. Although it is impossible for us to get the $h \rightarrow 0$ profile, we want to compare our data with their prediction in the limit that $\epsilon = 0$. To do so, recall that total curvature, as mentioned in Sec. I, is related to the shape through $C = [\psi(\theta)+\ddot{\psi}(\theta) - \psi(\theta) \dot{\psi}^2(\theta)/(1+\psi^2(\theta))]/r$. Since $\psi(\theta) \propto \epsilon$, we have $Cr/\epsilon = a_0 + a_1 \epsilon^2 + a_2 \epsilon^4 + O(\epsilon^6)$, where $a_0, a_1$ and $a_2$ could depend on $\theta$. Hence, from this equation, we can extrapolate the curvature at $\epsilon=0$ by fitting the curvature at three known values of $\epsilon$ with basis functions $\{1,~\epsilon^2,~\epsilon^4\}$. In FIG.~\ref{curvature}(b) we plot curvature profiles at $\epsilon =0.20,~0.15,~0.10$ and 0. The $\epsilon \rightarrow 0$ profile is extrapolated from other three profiles using the fitting theme discussed above. Solid line is the theory curve of exact solution. As we can see from this graph, as $\epsilon$ decreases, the curvature profile becomes closer to the theory curve that is based on the assumption that $\epsilon=0$ and $h=0$. Our $\epsilon=0$ profile agrees well with the theory curve in the central peak region. However, striking difference still exists in the range between the take-off positions and negative maxima. This discrepancy can be explained by the effect of nonzero thickness. For real sheet, $h \ne 0$, from Eq. (\ref{strain}), we observe that $\dot{\phi}$ tends to avoid jump at take-off positions in the non-contact region where external pressure $P=0$, otherwise there would be singularity in the strains. Therefore the observed curvature profile has to be rounded rather than having an abrupt kink in slope, which requires that curvature profile be curved up as the sheet leaves the container, just like what is displayed in the plot. Besides, the same reason can explain the slight difference near take-off points in the curvature profiles at three distances in FIG.~\ref{curvature}(a). For smaller $r$, the curvature may be more influenced by nonzero thickness effect, so the profile at $r=30a$ is more rounded than the profile at $r=50a$. In addition, we note that the theory curve based on the model proposed by Chaieb \textit{et al} \cite{chaieb1} does not match our data.

From the curvature file, we can measure the opening angle of $d$-cone. The opening angle is defined as the angular distance between the two ``take-off points'' where the sheet loses contact with the container. For asymptotically thin sheets with small deformation, this angle is predicted to be $138^{\mathrm{o}}$ from the exact solution, as illustrated by the solid line of FIG.~\ref{curvature}(b). As $\epsilon$ decreases to zero, the opening angle measured from curvature profiles tends to converge to that indicated by the solid line. We expect the agreement with the prediction when the elastic thickness $h$ vanishes. This value was approximately confirmed by experiments, which yielded $130^{\mathrm{o}}$ \cite{cerda-prl}.

\subsection{Normal Force}

\begin{figure}[!hbt]
\begin{center}
\includegraphics[width=0.45\textwidth]{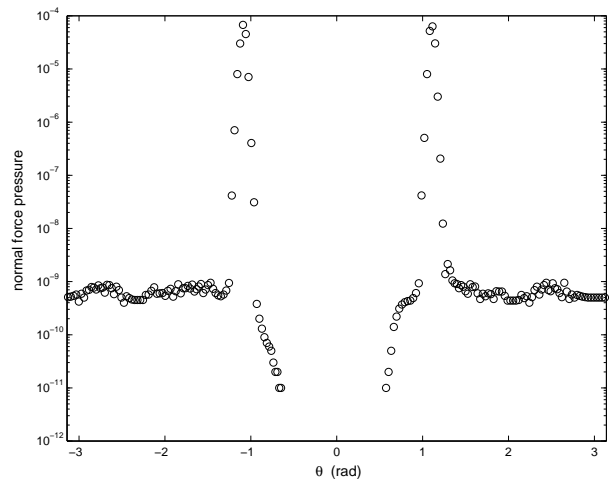}
\caption{Azimuthal profile of normal force pressure. This is for the shape shown in FIG.~\ref{shape}. Normal force pressure is measured in relative units. The right peak is located at 1.12 rad; the left peak is located at $-1.09$ rad. }
\label{normalforce}
\end{center}
\end{figure}

The azimuthal profile of the normal force pressure from the container is displayed in FIG.~\ref{normalforce}. It is evident that we do observe sharp peaks at the take-off positions, which confirm the $\delta$-function term in the normal force pressure proposed in \cite{maha-new}. The pressure drops quickly to zero as one enters the buckled region where the sheet bends away from container. The angular separation of the take-off positions is about 2.21 rad = $127^{\mathrm{o}}$. To verify that the sharp peaks in the normal force pressure have the proper strength, we calculate the ratio of the normal force from the $\delta$-function term to that from the $\theta$-independent term. The ratio is found to be $0.70$ from our data, compared well with the theoretical prediction $0.69$ obtained in Sec. II. 

Our measurements in Sec. III.B and this section confirm that our simulation accurately represents an elastic sheet as desired. We now report the observed behavior of the core region. 

\subsection{Scaling of Core Size}     

\begin{figure}[!hbt]
\begin{center}
\includegraphics[width=0.5\textwidth]{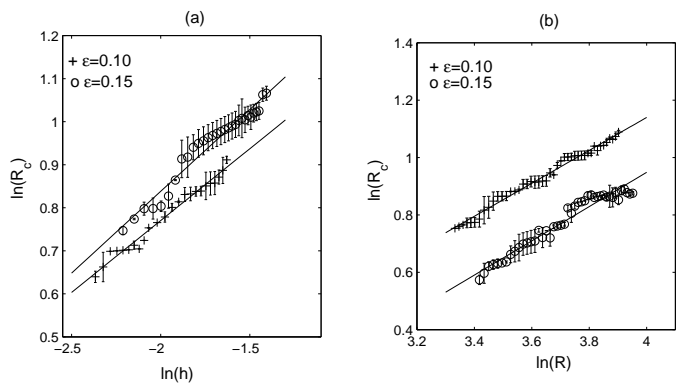}
\caption{The plots of (a) $R_c$ versus $h$ and (b) $R_c$ versus $R$ at $\epsilon=0.10$ and $\epsilon=0.15$. $R$ is fixed for plots in (a); $h$ is fixed for plots in (b). The straight lines are power law fits. The fitted values of power in (a) are $0.334 \pm 0.022$ for $\epsilon=0.10$ and $0.380 \pm 0.014$ for $\epsilon=0.15$. The fitted values of power in (b) are $0.574 \pm 0.021$ for $\epsilon=0.10$ and $0.597 \pm 0.032$ for $\epsilon=0.15$. The error bars in the graph come from the numerical hysteresis effect when thickness of sheet or radius of container is increased and decreased through the same value during the energy minimization process. }
\label{dcone_core}
\end{center}
\end{figure}

To study the scaling law of core region size, it is natural to start estimating core size by finding the radius of curvature of crescents. From azimuthal profiles of curvature at different distances, we find the locations of the triangles with maximal negative curvature on both sides at each fixed distance. Centers of these triangles are projected onto $x-y$ plane. The best fitted straight lines to the projection points on two sides serve as the asymptotic lines of a hyperbola, and the central forcing point is taken as the midpoint of the same hyperbola. This process is illustrated in FIG.~\ref{two_sketch}(b). $R_c$ is the radius of curvature at the midpoint of the fitted hyperbola shape. FIG.~\ref{dcone_core} shows the dependence of $R_c$ measured in this way on $h$ and $R$ when deformation $\epsilon$ is fixed at $0.10$ and $0.15$. We can observe power law dependence from these plots. The power law fits give $0.334 \pm 0.022 $ and $0.380 \pm 0.014$ for thickness dependence, and give $0.574 \pm 0.021$ and $0.597 \pm 0.032$ for radius dependence. These values roughly agree with the $1/3$ and $2/3$ exponents proposed in Ref. \cite{cerda-nature}.

As pointed out above, however, the lattice model can only accurately simulate an elastic sheet where the radius of curvature, $1/C$, is locally much greater than the lattice spacing. We expect this condition not to be well satisfied in the core region, where singularity happens. Indeed, our data indicate that $Ca$ can be as big as $0.5$ in this region even for small $\epsilon$. Thus we have no hope of maintaining complete accuracy in this region. In addition, we find that the curvature data do not exhibit the kind of shape as shown in FIG.~\ref{curvature} at small distances. Therefore, in the above procedure of measuring $R_c$ geometrically, we have to use curvature profiles at big distances, which makes it not so accurate to determine $R_c$ in this way since $R_c$ is supposed to be determined from the information in the neighbourhood of the tip. 

Our second approach of looking for the scaling relations is to use the force equation (\ref{forceeq}). In our simulation, we keep the spring constant $k$ fixed ($k=1$), which is equivalent to fixing two-dimensional Young's modulus ($Yh$) since we have $Yh=2ka/\sqrt{3}$. Hence $\kappa \propto Yh^3 \propto h^2$. From Eq. (\ref{forceeq}) we have 
\beq
F \approx \frac{h^2}{BR}\left(-p\ln{h}-q\ln{R}+(p+q)\ln{R_p} + D \right)~~,
\label{finalforce}
\eeq
where $B$ and $D$ only depend on $\epsilon$, which we now fix at $0.10$. Fixing $\epsilon$ is realized through a process of several minimizations. We slowly adjust the pushing force until the desired $\epsilon$ value is reached. In every step of the process, the previously obtained minimized configuration is used as the input for the next minimization procedure.

\begin{figure}[!hbt]
\begin{center}
\includegraphics[width=0.5\textwidth]{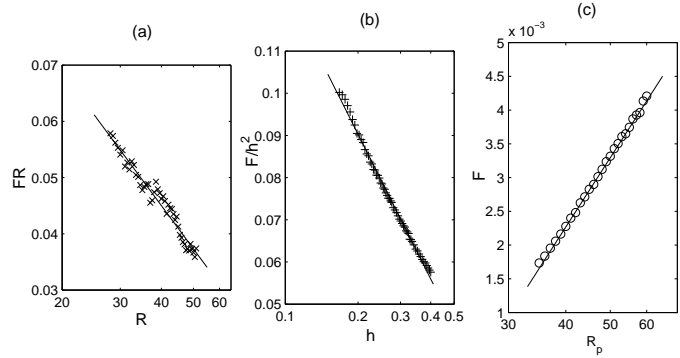}
\caption{Force plots when $\epsilon$ is fixed at $0.10$. (a) $FR$ versus $R$ semilog plot. The data is fitted to be $FR=-0.0342 \times \ln{R}+0.1714$. The parameters we use are $k=1$, $J=0.0013$, $R_p=60a$. (b) $F/h^2$ versus $h$ semilog plot. The data is fitted to be $F/h^2=-0.0494\times \ln{h} +0.0108$. The parameters we use are $k=1$, $R=37a$, $R_p=60a$. (c) $F$ versus $R_p$ semilog plot. The data is fitted to be $F=0.0046 \times \ln{R_p}-0.0147$. The parameters we use are $k=1$, $J=0.003$, $R=25a$. The relation between exponents $p$ and $q$ is $q=1.879p$.}
\label{force1}
\end{center}
\end{figure}

To find the relations between exponents, we first fix $h$ by fixing the bending coefficient $J$ defined in Sec. III.A, and measure force $F$ on the minimized shapes for different values of $R$. FIG.~\ref{force1}(a) gives the semilog plot of $FR$ versus $R$ when $h$ is fixed at $0.102a$ and $R_p$ is fixed at $60a$. The linear feature of the plot agrees with the prediction from Eq. (\ref{finalforce}). We denote the slope and intercept of the best fitted line by $S_1$ and $I_1$, respectively. Then $S_1=-0.0342 \pm 0.0011$ and $I_1=0.1714 \pm 0.0042 $. From Eq. (\ref{finalforce}) we find
\begin{subequations}
\bea
I_1 B & =  & h^2(-p\ln{h}+(p+q)\ln{R_p}+D) \label{eq1} \\
S_1 B & =  & h^2(-q)      \label{eq2}   
\eea
\end{subequations}
Next, for the same value of $\epsilon$, $R=37a$ and $R'_p=60a$ are fixed while $h$ is varied as the corresponding equilibrium force $F$ is measured. $F/h^2$ versus $h$ semilog plot is shown in FIG.~\ref{force1}(b). Similarly the slope of the best fitted line $S_2=-0.0494 \pm 0.0005$ and intercept $I_2=0.0108 \pm 0.0006$, which according to Eq.~(\ref{finalforce}) are given by
\begin{subequations}
\bea
I_2 B & = & \frac{1}{R}(-q \ln{R} +(p+q) \ln{R'_p} + D)  \label{eq3}  \\
S_2 B & = & \frac{1}{R}(-p)  \label{eq4}  
\eea
\end{subequations}
Notice that Eqs.~(\ref{eq1}), (\ref{eq2}), (\ref{eq3}) and (\ref{eq4}) constitute a homogeneous system of four linear equations in four variables $\{p, q, D, B\}$. In order to have non-trivial solutions, the determinant of its coefficient matrix must vanish. To check this condition, we write out the determinant of dimensionless matrix of coefficients explicitly \\
\begin{center} $\begin{vmatrix} (\ln{R_p}-\ln{h}) & \ln{R_p} & 1 & -I_1/h^2 \\ 0 & 1 & 0 & S_1/h^2 \\ \ln{R'_p} & (\ln{R'_p}-\ln{R}) & 1 & -I_2R \\ 1 & 0 & 0 & S_2R  \end{vmatrix} = (\ln{R}/h^2)S_1+I_1/h^2-(R\ln(hR'_p/R_p))S_2-RI_2$~~, \end{center} 
which is calculated to be $0.016 \pm 0.861$, indeed including 0. In addition, we obtain the relations between these variables: $q = (1.879 \pm 0.428) p$ and $B = (0.547 \pm 0.005) p$.

The third relationship we can extract from force equation (\ref{finalforce}) is $F$ versus $R_p$. To test this, we keep the $h_1=0.1549a$ and $R_1=25a$ fixed while measuring pushing forces for different $R_p$. The plot is shown in FIG.~\ref{force1}(c). On the one hand, the slope of the best fitted line is $0.00461 \pm 0.00003$ and intercept is $-0.0147 \pm 0.0001$. On the other hand, we can calculate the slope and intercept of this graph using the relations between $p$, $q$, $D$ and $B$ obtained above. Specifically, from Eq.~(\ref{finalforce}) we have slope $S_3=h_1^2(p+q)/(BR_1)=0.0051 \pm 0.0008$, and intercept $I_3=h_1^2/(BR_1)(-p\ln{h_1}-q\ln{R_1}+D)=-0.0157 \pm 0.0026$, both in good agreement with the direct fitted values from plot. These tests verify the self-consistency of our data and the validity of the form of the force equation, thus supporting the scaling behaviour of $R_c$.  

The useful information we obtain from these tests concerning the scaling exponents is the relation between $p$ and $q$. We can not determine the value of $\lambda$ in Eq. (\ref{forceeq}) just from these tests. However, geometrical measurements show that $R_c$ varies very little with $R_p$. Indeed, we find that the standard deviation of values of $R_c$ at different $R_p$ is only about a couple percents of the mean value of $R_c$. It seems this observation serves as a good footing for us to believe that $R_c$ is independent of $R_p$. If we make this assumption, that is, if we take $\lambda=0$, then from the relation between $p$ and $q$, we can obtain $p= 0.355 \pm 0.053$. This is close to the corresponding value from geometrical measurements. 

Besides $\epsilon=0.10$, we perform the above tests of the force equation for other values of $\epsilon$. The results are summarized in TABLE~\ref{results}. We notice that all the values of $p$ are consistent with scaling exponent of $1/3$. In addition, by comparison of Eq.~(\ref{forceeq}) and Eq.~(\ref{finalforce}), it is easy to see that $B$ is inversely proportional to $\epsilon$. This relation can be readily confirmed from the data shown in the table.  

As mentioned in Sec. II, $R_c$ scaling is expected to break for sufficiently large $R/h$. However, in our simulations discussed above, all the data are consistent with the assumption that there exists a scaling law of $R_c$, which indicates that the asymptotic regime of large $R/h$ is not yet reached. The values of $R/h$ in our simulations vary between 100 and 500. We conclude that the lower limit of this asymptotic regime should be at least above 500. It is unsettling that the asymptotic regime should require such large $R/h$, but similar behavior occurs in related phenomena. For example, the asymptotic scaling of the stretching ridge requires a ratio of $R/h$ of over a thousand \cite{witten}.

\begin{figure}[!hbt]
\begin{center}
\includegraphics[width=0.5\textwidth]{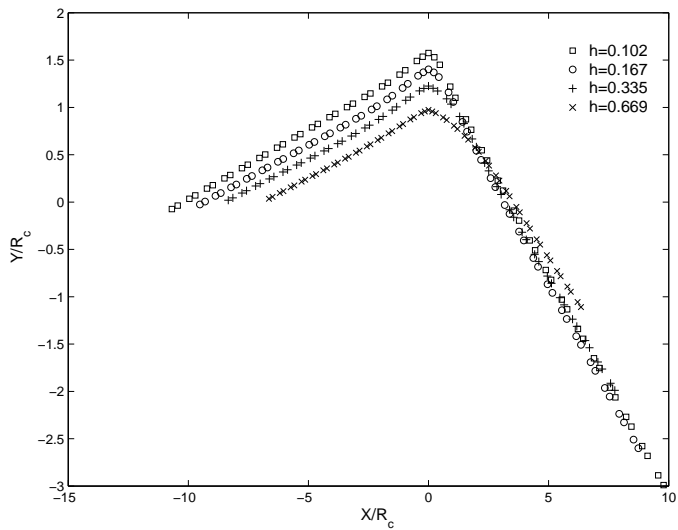}
\caption{Intersections of a $d$-cone with a vertical plane consisting of its maximally deflected line and $z$-axis, for four different thicknesses. Only the region near the peak is shown here. Both scales are normalized by the corresponding $R_c$ for values of $h$ as indicated in the legend. Thicknesses are in the units of lattice spacing $a$.}
\label{tran_project}
\end{center}
\end{figure}

The reason of this unusually high lower limit may be the following. The minor radius of the crescent is expected to be asymptotically much smaller than $R_c$. However, our observation shows that this minor radius is comparable with $R_c$ for the range of $R_c/h$ we are able to simulate. In this respect we see directly that our system is not asymptotic. We have noted in Sec. II that the system's asymptotic behavior depends strongly on how this minor radius behaves. Thus one cannot expect asymptotic scaling of $R_c$ without asymptotic behavior of the minor radius. Since our simulations do not reach this behavior, we should not be surprised that $R_c$ may not show the asymptotic scaling, either. To demonstrate that the minor radius is not asymptotic, we show its behavior in FIG.~\ref{tran_project}. This plot shows the intersection of the surface with a vertical plane passing transverse to the crescent. Thus the radius of curvature of this curve at its peak is the minor radius. Both the horizontal and vertical scales are normalized by $R_c$. It is evident from the plot that this normalized width is on the order of unity for all of the four thicknesses; this means the minor radius of crescent is comparable with $R_c$. In addition, the feature that the minor radius of crescent relative to $R_c$ is growing smaller as thickness decreases provides evidence that the observed scaling behavior of $R_c$ in our simulations is non-asymptotic.

\begin{table}[!hbt]
\begin{center}
\caption{\label{results}Results of testing force equation for different values of $\epsilon$}
\begin{ruledtabular}
\begin{tabular}{|c|c|c|c|}
$\epsilon$ & Determinant &  $B/p$ &  $p$  \\
\hline
$0.10$  &  $0.016 \pm 0.861$  & $0.547 \pm 0.005$ & $0.355 \pm 0.053$  \\
$0.15$  &  $-0.053 \pm 0.973$ & $0.367 \pm 0.003$ & $0.344 \pm 0.040$  \\
$0.20$  &  $0.056 \pm 2.076$  & $0.267 \pm 0.001$ & $0.381 \pm 0.071$  \\
$0.25$  &  $0.163 \pm 2.720$  & $0.203 \pm 0.001$ & $0.410 \pm 0.073$ \\
\end{tabular}
\end{ruledtabular}
\end{center}
\end{table}

\section{DISCUSSION}

In this paper we have explored properties of the conical singularity of a developable cone, especially the scaling of core region size. We have found out two types of stresses in the outer region of a single $d$-cone that scale differently with the distance to the tip. One type of stress arises from the normal force balance of the sheet without the external load, and scales as $1/r^2$. The other type of stress scaling as $1/r$ is needed to balance the external pushing force. However, it is revealed that the contribution of both stresses to the stretching energy is negligible compared with bending energy of the outer region. 
    
We also examined the normal force pressure from the container. The jump in the curvature derivative $\dot{\phi}/r$ requires a $\delta$-function pressure from the container edge at the take-off points. We verified that this singular pressure is present with the predicted magnitude \cite{maha-new}. As a consequence, a substantial fraction of the container forces comes from this $\delta$-function term. 

Our numerical tests of pushing force equation suggest the existence of scaling behavior of $R_c$ in the regime we studied, that is, when $R$ is comparable with $R_p$. We obtain a simple proportionality factor between exponents $p$ and $q$. Since $R_c$ has dependence on $h$, i.e. $p$ is nonzero, it follows that $q$ is nonzero, which implies $R_c$ must have a dependence on $R$. This is somewhat counter-intuitive. Moreover, geometrical measurements of core size provide suggestive numerical evidence that $R_c$ is independent of $R_p$. By taking this assumption, we are led to a scaling law suggesting $R_c \sim h^{1/3}R^{2/3}$. Although our numerical results are consistent with the scaling prediction of Ref. \cite{cerda-nature}, we were unable to justify the arguments leading to their prediction. The factors determining $R_c$ are necessarily subtle, since the dominant energy, $B_0$, depends only logarithmically on $R_c$. Until a clear justification for this scaling can be found and validated, the apparent scaling we observed must be viewed as provisional. 

Our work in progress aims to explore the energetics leading to $R_c$ scaling in more details. Our preliminary findings suggest new features that may help to resolve this issue. First, one may construct variants of a $d$-cone that require no forcing at the core. Our preliminary data show that these variants have qualitatively different scaling behaviour than the conventional $d$-cones. Second, conventional $d$-cones appear to obey an unanticipated constraint at the container edge. The mean curvature appears to vanish there for a wide range of $d$-cone shapes. We anticipate that the energy focusing in $d$-cones will prove to be rich and revealing.

\begin{acknowledgments}

The authors would like to thank Enrique Cerda and Sergio Rica for enlightening discussions. This work was supported in part by the National Science Foundation its MRSEC Program under Award Number DMR-0213745. 

\end{acknowledgments}

\end{document}